\documentclass[preprint,12pt]{elsarticle}

\usepackage{amssymb}

\usepackage{lineno}

\journal{Chaos Soliton Fract.}

\begin{document}

\newcommand{\bra}{\left\langle}
\newcommand{\ket}{\right\rangle}
\newcommand{\tbox}[1]{\mbox{\tiny #1}}

\begin{frontmatter}



\title{Spectral and localization properties of random bipartite graphs}


\author{C. T. Mart\'{\i}nez-Mart\'{\i}nez}
\address{Instituto de F\'isica, Benem\'erita Universidad Aut\'onoma de Puebla,
Apartado Postal J-48, Puebla 72570, Mexico}
\address{Institute for Biocomputation and Physics of Complex Systems (BIFI), University of Zaragoza, 50018 Zaragoza, Spain}

\author{J. A. M\'endez-Berm\'udez}
\address{Instituto de F\'isica, Benem\'erita Universidad Aut\'onoma de Puebla, 
Apartado Postal J-48, Puebla 72570, Mexico}
\address{Departamento de Matem\'{a}tica Aplicada e Estat\'{i}stica, Instituto de Ci\^{e}ncias Matem\'{a}ticas e de Computa\c{c}\~{a}o, Universidade de S\~{a}o Paulo - Campus de S\~{a}o Carlos, Caixa Postal 668, 13560-970 S\~{a}o Carlos, SP, Brazil}

\author{Yamir Moreno}
\address{Institute for Biocomputation and Physics of Complex Systems (BIFI), University of Zaragoza, 50018 Zaragoza, Spain}
\address{Department of Theoretical Physics, University of Zaragoza, 50009 Zaragoza, Spain}
\address{ISI Foundation, Turin, Italy} 

\author{Jair J. Pineda-Pineda}
\address{Regional Development Sciences Center, Autonomous University of Guerrero, 
Los Pinos s/n, Suburb El Roble, Acapulco, Guerrero, Mexico}

\author{Jos\'e M. Sigarreta}
\address{Facultad de Matem\'aticas, Universidad Aut\'onoma de Guerrero, 
Carlos E. Adame No.54 Col. Garita, Acalpulco Gro. 39650, Mexico}

\begin{abstract}
Bipartite graphs are often found to represent the connectivity between the components of many systems such as ecosystems. A bipartite graph is a set of $n$ nodes that is decomposed into two disjoint subsets, having $m$ and $n-m$ vertices each, such that there are no adjacent vertices within the same set. The connectivity between both sets, which is the relevant quantity in terms of connections, can be quantified by a parameter $\alpha\in[0,1]$ that equals the ratio of existent adjacent pairs over the total number of possible adjacent pairs. Here, we study the spectral and localization properties of such random bipartite graphs. Specifically, within a Random Matrix Theory (RMT) approach, we identify a scaling parameter $\xi\equiv\xi(n,m,\alpha)$ that fixes the localization properties of the eigenvectors of the adjacency matrices of random bipartite graphs. We also show that, when $\xi<1/10$ ($\xi>10$) the eigenvectors are localized (extended), whereas the localization--to--delocalization transition occurs in the interval $1/10<\xi<10$. Finally, given the potential applications of our findings, we round off the study by demonstrating that for fixed $\xi$, the spectral properties of our graph model are also universal. 
\end{abstract}

\begin{keyword}
Bipartite graphs \sep delocalization transition \sep spectral properties
\PACS 64.60.aq \sep 89.75.Da \sep 05.45.Mt \sep 73.20.Jc
\end{keyword}
\end{frontmatter}



\section{Introduction}

The latest developments in network science have largely contributed to a better understanding of the structure and dynamics of many real-wold complex systems \cite{Barrat08:book,Newman010:book,Boccaletti06:PR}. As a matter of fact, research done during the last 20 years have allowed to take key steps in our comprehension of seemingly diverse phenomena such as the large-scale spreading of diseases \cite{Vespignani2015,Arruda2018}, information dissemination \cite{Newman010:book}, cascading failures \cite{stanley2010}, diffusion dynamics \cite{Gomez2013, tejedor2018,masuda2017} and more recently, on how multilayer systems work \cite{Kivela2014,boccaletti2014,Aleta2019}. These advances are not only at a theoretical level. The increasing availability of new and rich data as well as our computational capabilities have made it possible to move from studying synthetic models, to characterize and model realistic systems. 

During these years, networks have been studied from many different angles, ranging from more theoretically-grounded studies (in the best tradition of graph theory) to fully data-driven models. Sometimes, the architecture of the substrate network is known and thus, it could be modeled explicitly. However, it is often the case in which the networks are synthetic either because we do not know the real connection patterns or because we need to simplify the structure of the system to enable analytical approximations. In the latter scenario, one reasonable assumption is to generate random graphs, so that one gets rid of possible correlations and isolates the impact of the connectivity among the system's constituents on its dynamics. Besides, random versions are often very useful as null models, that allow to individuate which properties of the system are truly unexpected and which are not \cite{cimini2019,payrato2019}.
 
Among the many results that can be highlighted, perhaps the most useful ones are those that relate the structure of networks with their dynamics through the analysis of the spectral properties of the adjacency or Laplacian matrices of such networks. For instance, it has been shown that it is possible to characterize the critical properties of a disease spreading process in terms of the largest eigenvalue of the adjacency matrix of the network on top of which the dynamics takes place \cite{Vespignani2015,Arruda2018}. Admittedly, the fact that the epidemic threshold, i.e., the point beyond which the system experiences a macroscopic outbreak, can be expressed in terms of topological properties makes it possible to study what are the effects of the topology on the dynamics of complex networked systems. Another important example of the previous relationship between structure and dynamics is given by synchronization phenomena, where one finds that the stability of a fully synchronized system can be studied in terms of the spectral properties of the substrate network \cite{Barrat08:book,Newman010:book,Boccaletti06:PR}.

In this paper, we follow the line of research mentioned above and study a class of networks that is often found in natural and artificial systems, namely, bipartite graphs. Within the classes of networks that have been analyzed in the last two decades, bipartite graphs have gone unnoticed in many regards, for instance, in relation to their spectral properties. We intend to fill this gap by studying the localization and spectral properties of random bipartite graphs within RMT approaches. This viewpoint has been successfully used to study some topological~\cite{MMRS19}, 
spectral~\cite{MAM15,MM19,GAC18}, 
eigenvector~\cite{MAM15,MM19}, and transport~\cite{MAM13} properties of ER--type random 
networks with a special focus on universality. Moreover, we have also performed scaling studies on other random network models, such as multilayer and multiplex networks~\cite{MFR17,MFR17b} and random--geometric and random--rectangular graphs~\cite{AMG18}. 

The rest of the paper is organized as follows. In Sec.~\ref{model} we define the random bipartite graph model we shall use in our study. 
Then, in Sec.~\ref{entropy} we perform a scaling analysis of the eigenvector properties (characterized 
by the Shannon or information entropy) of our bipartite graph model. The scaling analysis allows to 
define a universal parameter of the model that we validate in Sec.~\ref{spectra} with the scaling of 
the spectral properties (characterized by the distribution of ratios of consecutive energy-level spacings).
We summarize our results in Sec.~\ref{conclusions} also discussing possible applications within the domain of ecosystems and their stability.

\section{Bipartite graph model}
\label{model}

\begin{figure*}[t!]
\centering
\includegraphics[width=0.9\textwidth]{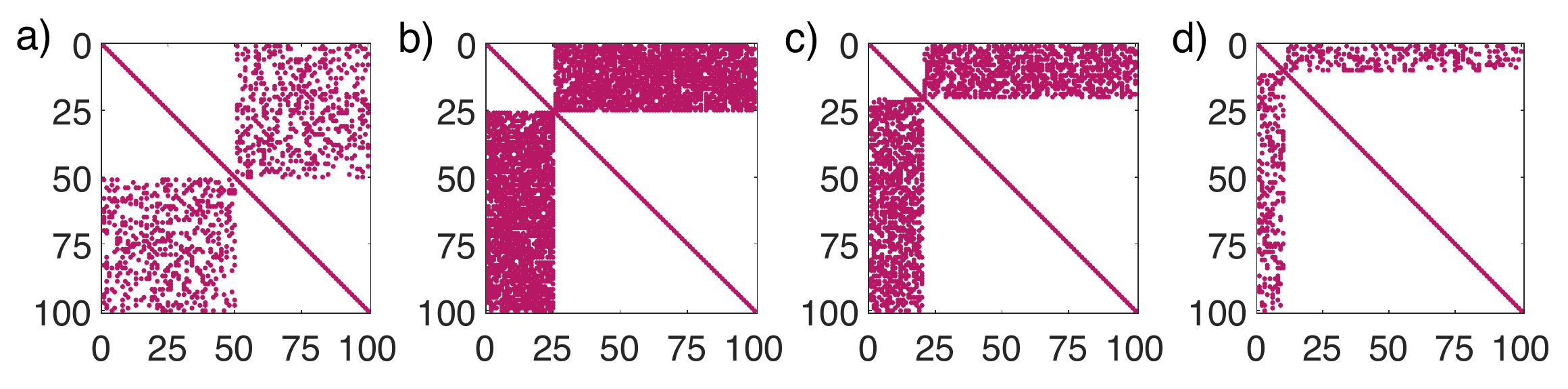}
\caption{Nonzero adjacency matrix elements of random bipartite graphs for some combinations of 
$m$ and $\alpha$: (a) $m=n/2$ and $\alpha =0.2$, (b) $m=n/4$ and $\alpha =0.75$, (c) $m=n/5$ 
and $\alpha =0.5$, (d) $m=n/10$ and $\alpha =0.25$. In all cases $n=100$.}
\label{Fig1}
\end{figure*}

We consider bipartite graphs composed by two disjoint sets with $m$ and $n-m$ vertices each such that 
there are no adjacent vertices within the same set, being $n$ the total number of vertices in the bipartite 
graph. The connectivity between both sets is quantified by the parameter $\alpha$ which is the ratio of 
current adjacent pairs over the total number of possible adjacent pairs; that is, vertices are isolated
when $\alpha=0$, whereas the bipartite graph is complete for $\alpha=1$. Vertices are connected randomly. We add to our bipartite graph model self-edges and further consider all edges to have random strengths, which allows that our bipartite graph model becomes a RMT model. Therefore, we define the corresponding adjacency matrices as members of the ensemble of $n\times n$ sparse real symmetric matrices whose non-vanishing elements are statistically independent random variables drawn from a normal distribution with zero mean $\bra A_{ij} \ket=0$ and variance $\bra |A_{ij}|^2 \ket=(1+\delta_{ij})/2$. According to this definition, a diagonal adjacency random matrix is 
obtained for $\alpha=0$, which is known as the Poisson ensemble in RMT terms. In Fig.~\ref{Fig1}, we show examples of adjacency matrices of random bipartite graphs with $n=100$ vertices and some combinations of $m$ and $\alpha$. Note that when labeling the vertices according to the set they belong to, the adjacency matrices of bipartite graphs have a block structure. 

Here we define $m$ (resp. $n-m$) as the number of vertices of the smaller (bigger) set. In this respect, the 
case $m=n/2$ is a limiting case where both sets have the same number of vertices, $m=n-m$. 
Moreover, the case $m=1$ is another limiting case in which the smaller set consists of a single vertex.
Thus, in what follows we will consider random bipartite graphs characterized by the parameter set $(n,m,\alpha)$ 
with $1\le m\le n/2$ and $0\le \alpha \le 1$. Notice that the case $m>n/2$ is redundant because it is equivalent to the interchange of the sets.

\begin{figure*}[t!]
\centering
\includegraphics[width=0.9\textwidth]{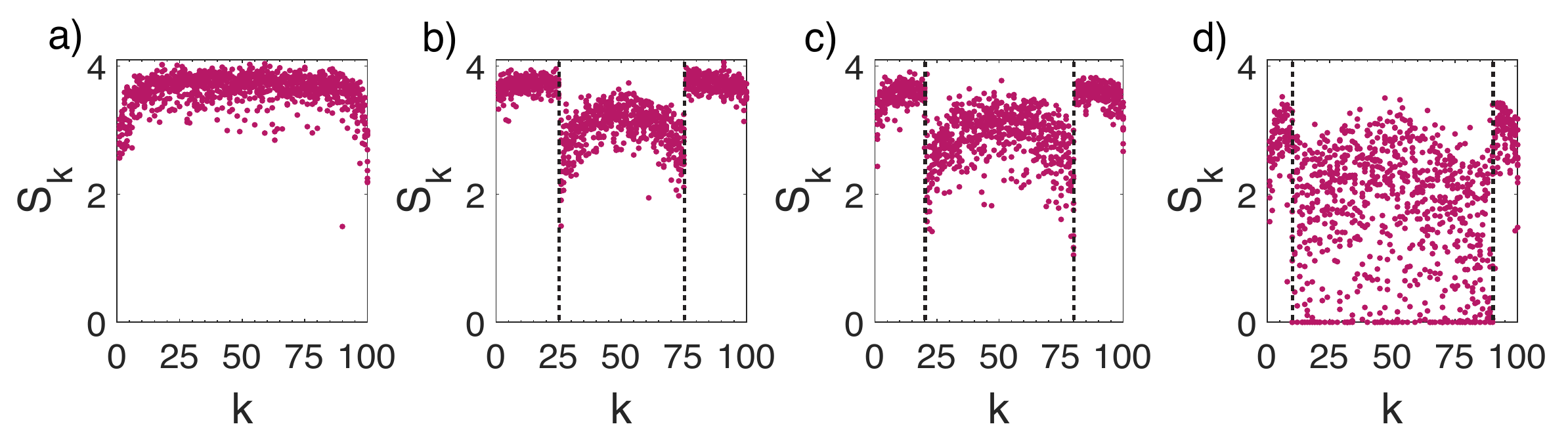}
\caption{Shannon entropies $S^k$ of the eigenvectors of ten realizations of the adjacency matrices 
shown in Fig.~\ref{Fig1}. Dashed lines in panels (b-d) separate groups of entropies characterized 
by different average values.}
\label{Fig2}
\end{figure*}

\section{Eigenvector properties. Scaling and universality}
\label{entropy}

In this study, we characterize the eigenvectors of random bipartite graphs by using information or 
Shannon entropy, which for the eigenvector $\Psi^k$ is given as
\begin{equation}
\label{S}
S^k = -\sum_{j=1}^n \left| \Psi^k_j \right|^2 \ln \left| \Psi^k_j \right| ^2 \ .
\end{equation}
$S^k$ measures the number of principal components of the eigenvector $\Psi^k$ in a given basis. Therefore, the latter quantity is a
good measure of eigenvector localization/delocalization. In fact, this
quantity has already been used to characterize quantitatively the complexity and localization properties of the 
eigenvectors of the adjacency matrices of several random network models (see 
examples in~\cite{MAM15,MM19,MFR17,MFR17b,AMG18} and references therein). Below we use exact numerical diagonalization to compute the eigenvectors $\Psi^k$ and eigenvalues $\lambda_k$ ($k=1\ldots n$) of the adjacency matrices of large ensembles of random bipartite graphs 
characterized by the parameter set $(n,m,\alpha)$. 

In Fig.~\ref{Fig2}, we present the Shannon entropies $S^k$ of the eigenvectors of ten realizations 
of the adjacency matrices shown in Fig.~\ref{Fig1}. Note that for $m=n/2$ all rows of the adjacency matrix have the same average number of nonzero off-diagonal elements, see Fig.~\ref{Fig1}(a), therefore the corresponding eigenvectors are expected to be equivalent
and they should have similar entropies; this can be verified in Fig.~\ref{Fig2}(a).
In contrast, for any $m<n/2$, $m$ rows of the adjacency matrix have a larger number of nonzero 
off-diagonal elements than the remaining $n-m$ rows, see Figs.~\ref{Fig1}(b-d). Hence, as it can be seen in 
Figs.~\ref{Fig2}(b-d), the entropies of the corresponding eigenvectors can be grouped into two sets 
characterized by different average values $\bra S \ket$ (see the dashed lines in these panels, which separate the two sets having different averages). Despite these differences, taking into account that we want to use the average entropy to find scaling properties in random bipartite graphs, and that for this purpose we need a single quantity regardless of the specific graph, we compute averages over all available eigenvectors, thus taking into account the contribution of both eigenvector sets.

From definition~(\ref{S}), it follows that $\bra S \ket=0$ when $\alpha=0$, since the eigenvectors of the (diagonal) adjacency matrices 
of our random bipartite graph model have only one non-vanishing component with magnitude equal to 
one. On the other hand, for $\alpha=1$ the bipartite graph is complete and $\bra S \ket$
gets its maximal value, $S_{\tbox{MAX}}$, for a given combination of $n$ and $m$. Thus, when $0<\alpha<1$ we
should observe $0<\bra S \ket<S_{\tbox{MAX}}$.

\begin{figure}[t]
\centering
\includegraphics[width=0.6\columnwidth]{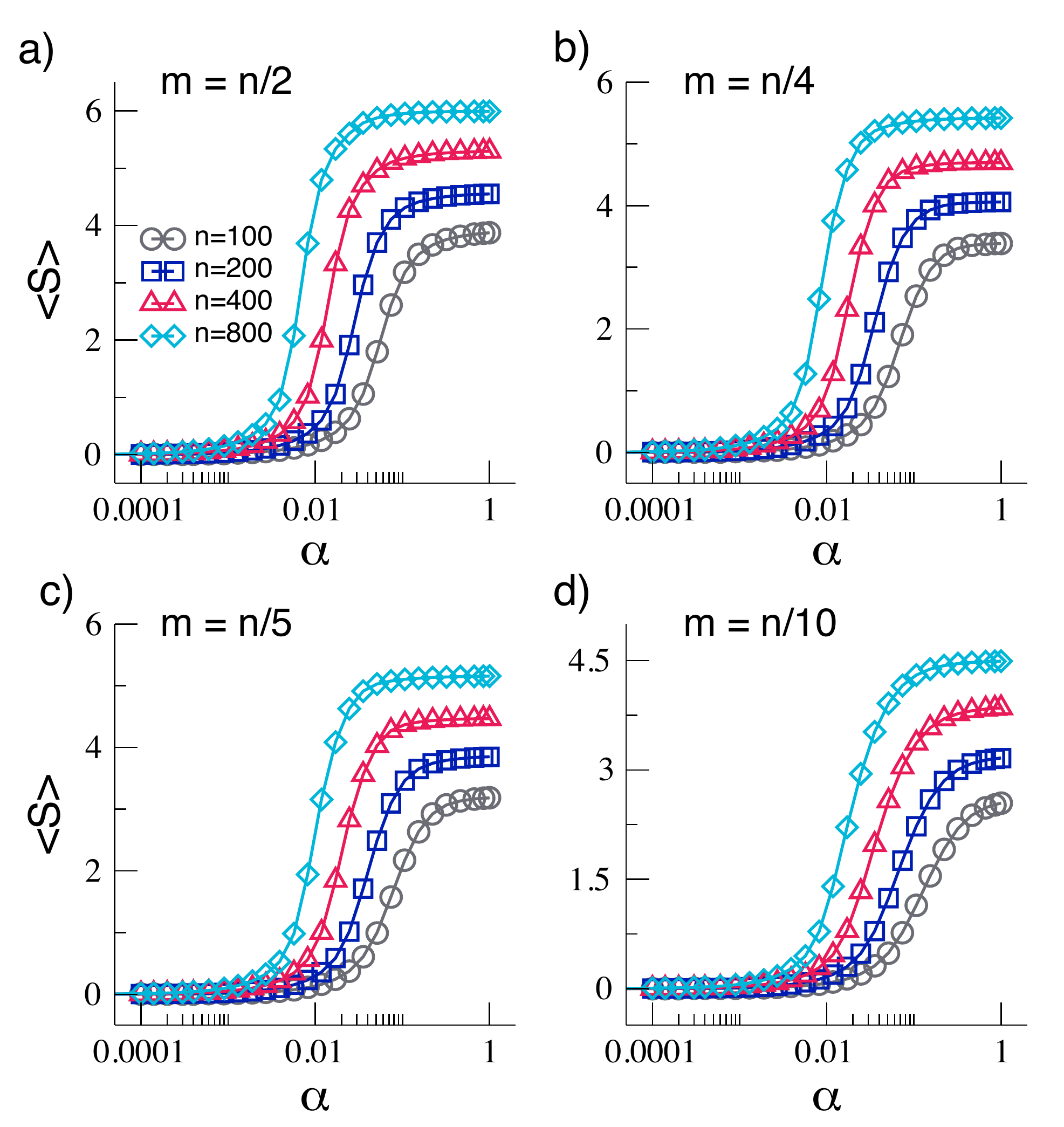}
\caption{Average Shannon entropy $\bra S \ket$ as a function of the connectivity $\alpha$ for random 
bipartite graphs (of sizes ranging from $n=100$ to 800) for several values of $m$ (as indicated in the panels).
Each symbol was computed by averaging over $10^6$ eigenvectors.}
\label{Fig3}
\end{figure}

\begin{figure}[h!]
\centering
\includegraphics[width=0.55\columnwidth]{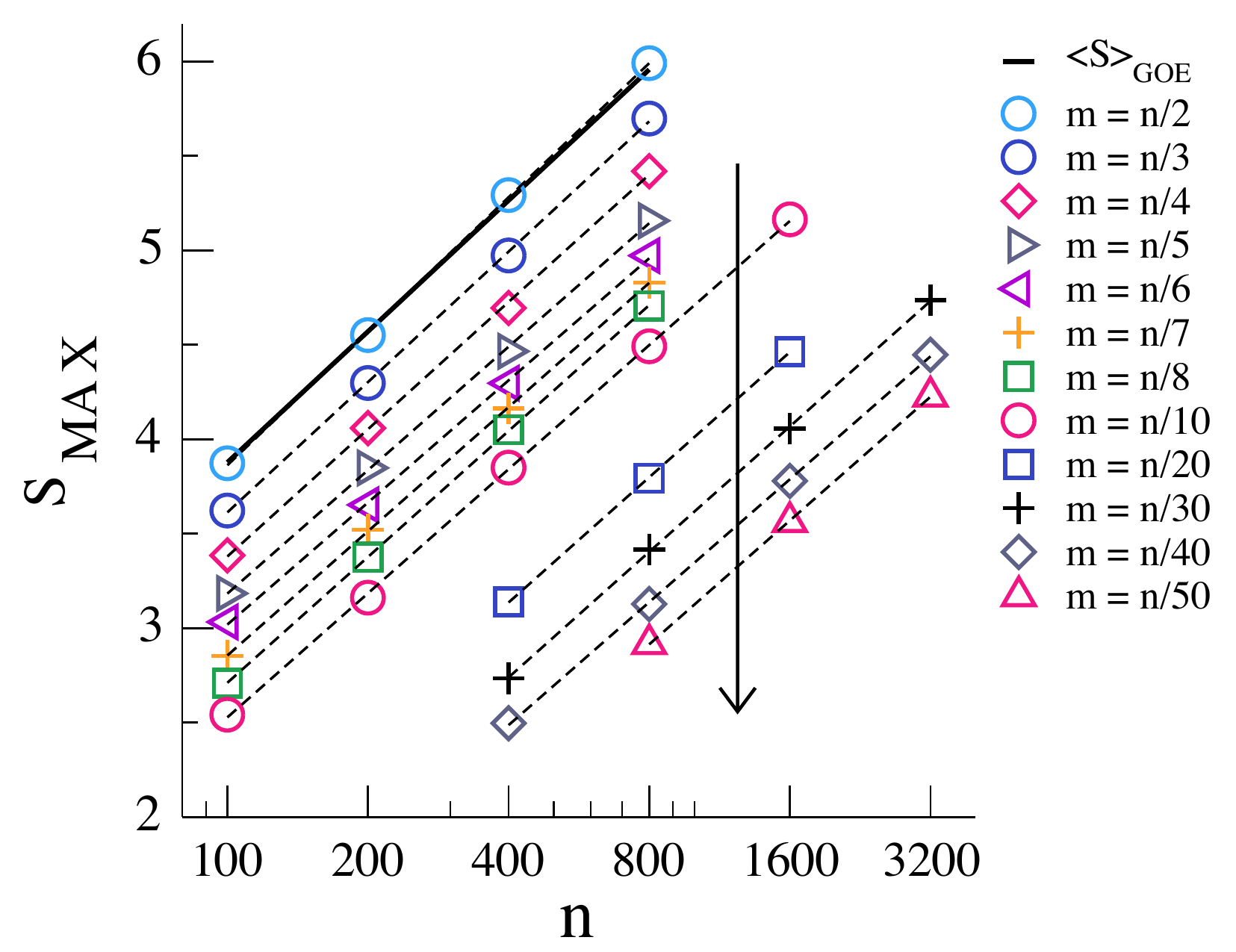} 
\caption{Maximum values of the Shannon entropy $S_{\tbox{MAX}}$ as a function of the bipartite graph size 
$n$ for several values of $m$. The thick black line corresponds to $\ln(n/2.07)$, the approximate value 
of $\bra S \ket_{\tbox{GOE}}$. The arrow indicates decreasing $m$.}
\label{Fig4}
\end{figure}

In Fig.~\ref{Fig3} we present the average Shannon entropy $\bra S \ket$ as a function of the connectivity 
parameter $\alpha$ for the eigenvectors of random bipartite graphs and for several parameter combinations. 
We observe that the curves of $\bra S \ket$, for any combination of $n$ and $m$, have a very similar 
functional form as a function of $\alpha$: The curves $\bra S \ket$ show a smooth transition from 
approximately zero to $S_{\tbox{MAX}}$ when $\alpha$ increases from $\alpha\sim 0$ (mostly isolated 
vertices) to one (complete bipartite graphs). Recall that when $\bra S \ket \approx 0$ the corresponding 
eigenvectors are localized (i.e., ~$\bra S \ket \approx 0$ defines the localized regime). In contrast, when
$\bra S \ket \approx S_{\tbox{MAX}}$, the corresponding eigenvectors are delocalized. Thus, the curves
of $\bra S \ket$ versus $\alpha$ in Fig.~\ref{Fig3} display the delocalization transition of the eigenvectors
of our random bipartite model. As a complementary information, in Fig.~\ref{Fig4} we report 
$S_{\tbox{MAX}}$, i.e., ~the value of $\bra S \ket$ at $\alpha=1$, of random bipartite graphs for several 
combinations of $n$ and $m$.

\begin{figure}[t]
\centering
\includegraphics[width=0.6\columnwidth]{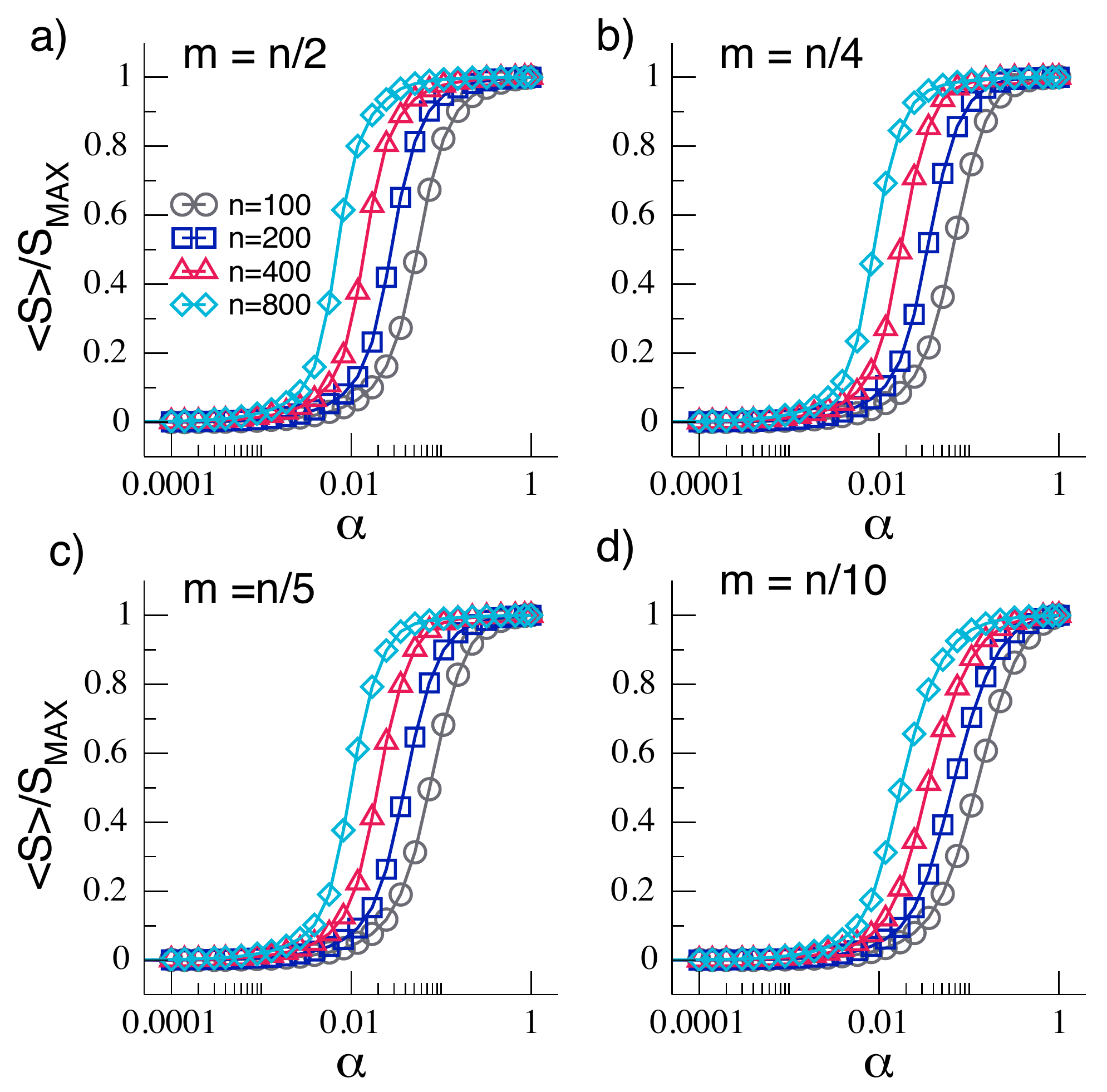}
\caption{Average information entropy $\bra S \ket$ normalized to $S_{\tbox{MAX}}$ as a function of the 
connectivity $\alpha$. Same data of Fig.~\ref{Fig3}.}
\label{Fig5}
\end{figure}
\begin{figure}[t]
\centering
\includegraphics[width=0.55\columnwidth]{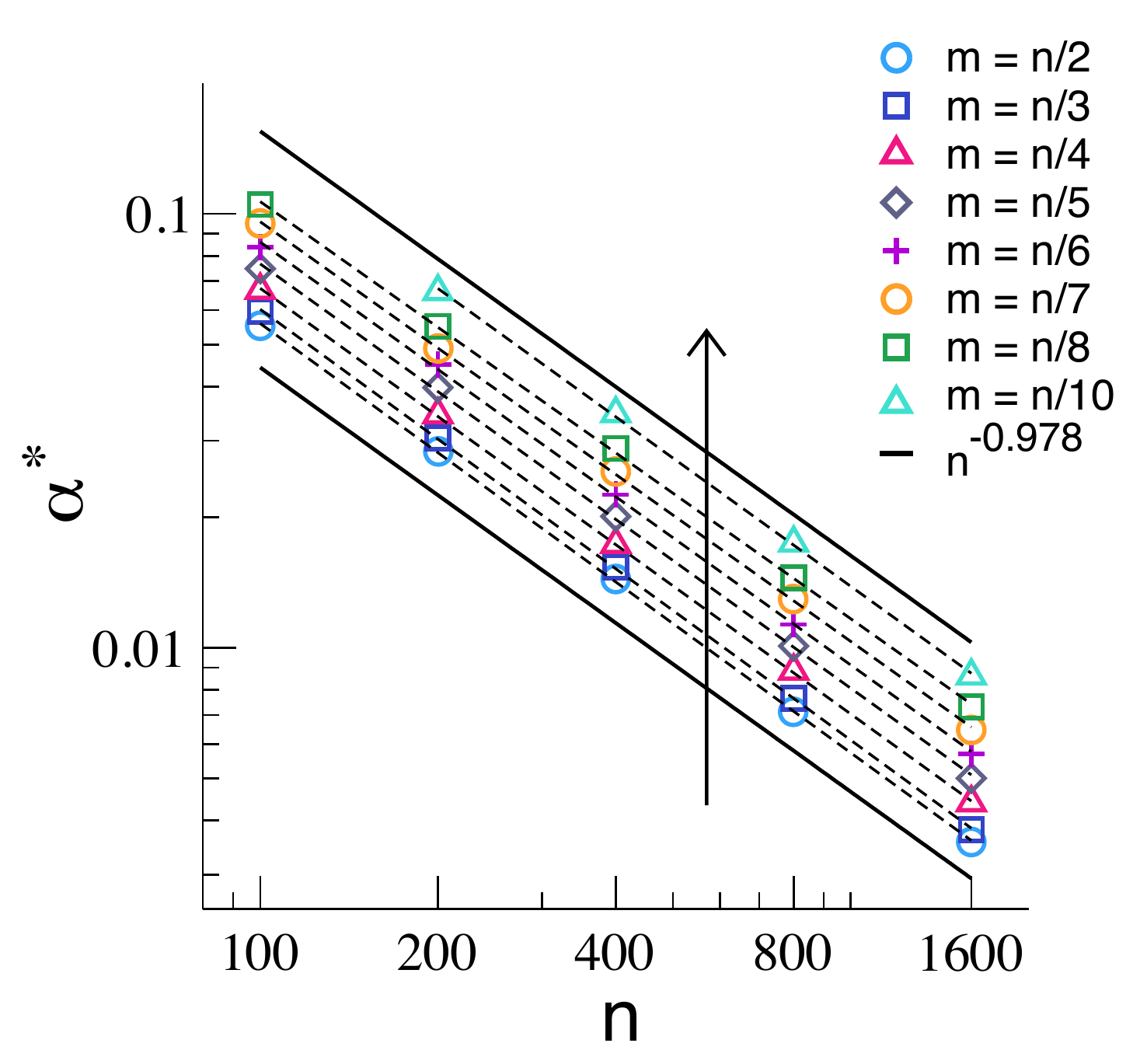}
\caption{Localization--to--delocalization transition point $\alpha^*$ (defined as the value of $\alpha$ 
for which $\bra S \ket/S_{\tbox{MAX}} \approx 0.5$) as a function of the bipartite graph size $n$ for several 
values of $m$. Dashed lines are the fittings of the data with Eq.~(\ref{scalingEq1}). The arrow indicates 
decreasing $m$.}
\label{Fig6}
\end{figure}
\begin{figure}[t]
\centering
\includegraphics[width=0.6\columnwidth]{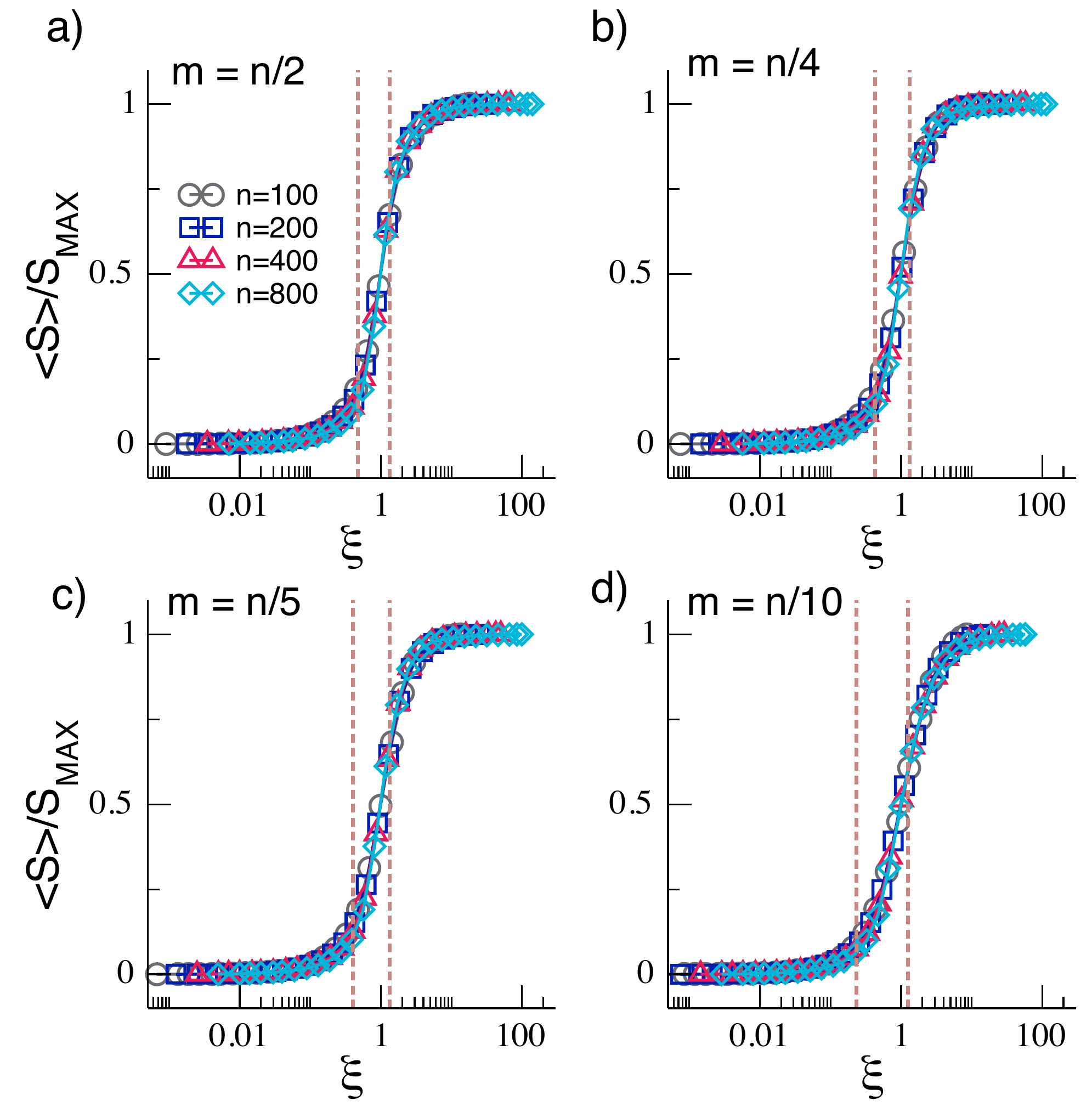}
\caption{Average information entropy $\bra S \ket$ normalized to $S_{\tbox{MAX}}$ as a function of the 
scaling parameter $\xi$, see Eq.~(\ref{xi}). Same data of Fig.~\ref{Fig3}. Dashed vertical lines indicate
the width of the transition region $\Delta$ defined as the full width at half maximum of the functions 
$d\bra S \ket/d\xi$ vs.~$\xi$.}
\label{Fig7}
\end{figure}
\begin{figure}[t]
\centering
\includegraphics[width=0.55\columnwidth]{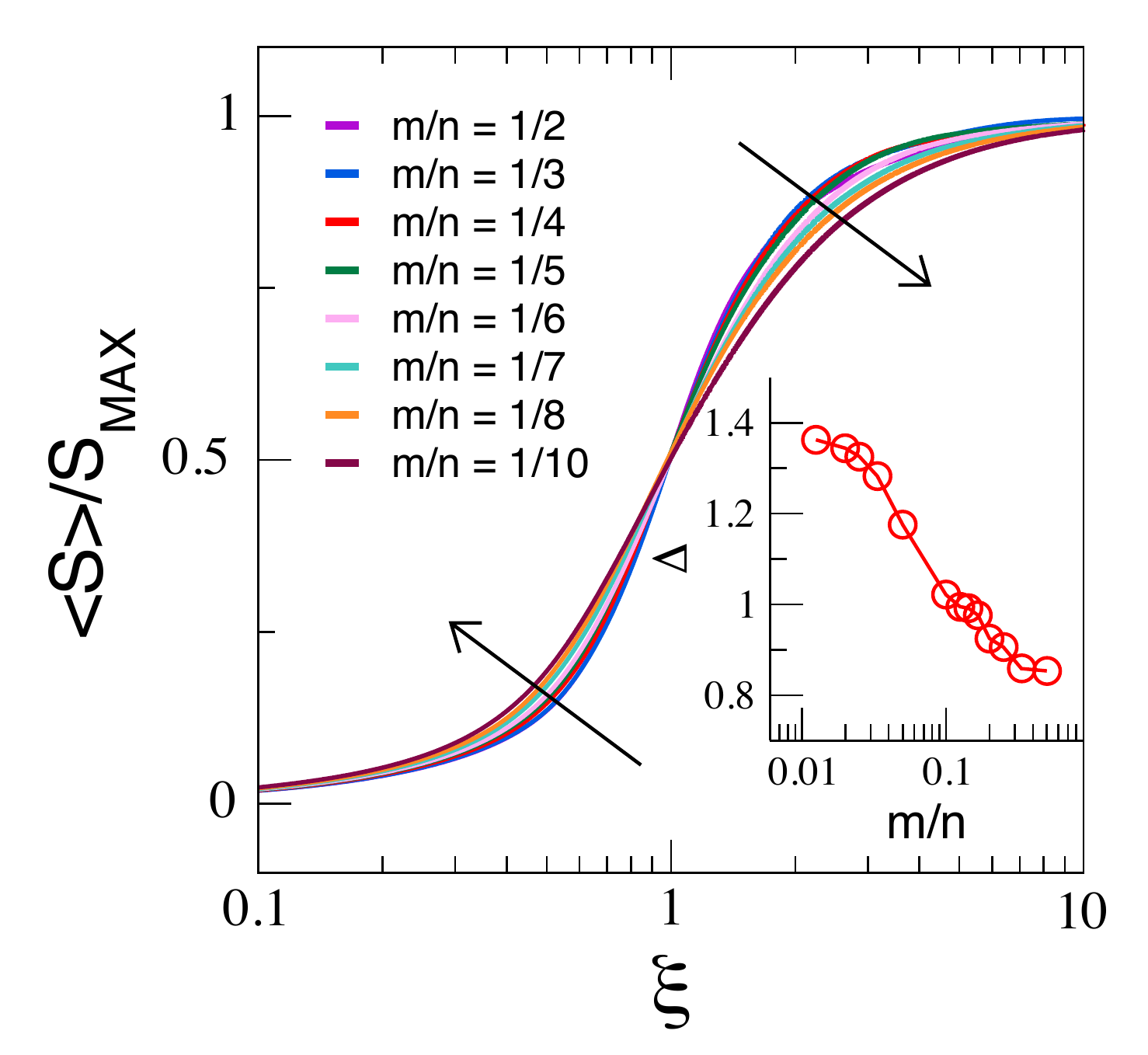}
\caption{Scaled curves for the Shannon entropy for random bipartite graphs with several values of $m/n$.
Arrows indicate decreasing $m/n$. All curves correspond to interpolated data with $n=800$. Inset:
Width of the transition region $\Delta$ as a function of $m/n$.}
\label{Fig8}
\end{figure}

It is important to stress that in our graph model with fixed $n$ the maximal number of nonzero adjacency 
matrix elements is obtained when $\alpha=1$ and $m=n/2$, but still in this case half of the off-diagonal 
adjacency matrix elements are equal to zero. Therefore the adjacency matrices of our random bipartite 
graphs never reproduce the Gaussian Orthogonal Ensemble (GOE) of RMT $-$the GOE is a random matrix 
ensemble formed by real symmetric random matrices $\bf A$ whose entries are statistically independent 
random variables drawn from a normal distribution with zero mean and variance 
$\bra |A_{ij}|^2\ket=(1+\delta_{ij})/2$, see e.g.~\cite{metha}. Accordingly, one should expect $S_{\tbox{MAX}}<\bra S \ket_{\tbox{GOE}}$, where 
$\bra S \ket_{\tbox{GOE}}\approx\ln(n/2.07)$ is the average entropy of the (random and delocalized) 
eigenvectors of the GOE. However, surprisingly, we observe that $S_{\tbox{MAX}} \approx \bra S \ket_{\tbox{GOE}}$
for $m=n/2$, while $S_{\tbox{MAX}}<\bra S \ket_{\tbox{GOE}}$ indeed occurs for any $m<n/2$, 
see Fig.~\ref{Fig4}. Also, from Fig.~\ref{Fig4}, we can clearly see that 
\begin{equation}
\label{SMAX}
S_{\tbox{MAX}}\propto\ln(n) \ .
\end{equation}
Therefore, we can conclude that the maximal entropy setup in our random bipartite graph model
corresponds to $m=n/2$ and $\alpha=1$ for which GOE statistics is observed for $\bra S \ket$ and
expected for other quantities.

Now, to ease our analysis, in Fig.~\ref{Fig5} we plot again $\bra S \ket$ but normalized to $S_{\tbox{MAX}}$.
The fact that these curves, plotted in semi-log scale, are just shifted to the left on the $\alpha$-axis 
when increasing $n$ makes it possible to hypothesize the existence of a scaling parameter that depends on $n$. In order to 
check this hypothesis and find such a scaling parameter, we first define a quantity that allows characterizing the position of the curves 
$\bra S \ket/S_{\tbox{MAX}}$ on the $\alpha$-axis: We choose the value of $\alpha$, that we label as 
$\alpha^*$, for which $\bra S \ket/S_{\tbox{MAX}} \approx 0.5$. Notice that $\alpha^*$ characterizes the 
localization--to--delocalization transition of the eigenvectors of our graph model. 

Figure~\ref{Fig6} shows the localization--to--delocalization transition point $\alpha^*$ as a function 
of $n$ for several values of $m$. The linear trend of the data (in log-log scale) in Fig.~\ref{Fig6} implies a 
power-law relation of the form
\begin{equation}
\label{scalingEq1}
\alpha^* = \mathcal{C} n^\delta \ .
\end{equation}
In fact, Eq.~(\ref{scalingEq1}) provides very good fittings to the data. The values of $\delta$ from the 
fittings are very close to -0.978 for all the values of $m$ considered here (see thick full lines in Fig.~\ref{Fig6}).
From this observation we can propose the following scaling for the curves $\bra S \ket/S_{\tbox{MAX}}$ vs~$\alpha$:
By plotting again the curves of $\bra S \ket/S_{\tbox{MAX}}$ now as a function of $\xi$, that we define as the 
ratio between the connectivity parameter and the localization--to--delocalization transition point
\begin{equation}
\label{xi}
\xi = \frac{\alpha}{\alpha^*} \propto \frac{\alpha}{n^\delta} \approx \alpha n^{0.978},
\end{equation}
we observe that curves for different bipartite graph sizes $n$ collapse on top of a single curve, see Fig.~\ref{Fig7}. That is, we conclude that, for a given ratio $m/n$, $\xi$ fixes the localization properties of the eigenvectors of the adjacency matrices of the random bipartite graphs, such that, when $\xi<1/10$ [$10<\xi$] the eigenvectors are localized [extended], while the localization--to--delocalization 
transition occurs in the interval $1/10<\xi<10$.

Even though we were able to scale the Shannon entropy curves for random bipartite graphs, as shown in
Fig.~\ref{Fig7}, there is still a dependence of those {\it universal} curves on the ratio $m/n$. 
To clearly show this, in Fig.~\ref{Fig8} we report
scaled curves of the Shannon entropy for several values of $m/n$ in the localization--to--delocalization transition 
region. Here we can observe that the larger the ratio $m/n$, the sharper the localization--to--delocalization 
transition. Thus, we characterize the width of the transition region, that we call $\Delta$, as the full width at 
half maximum of the functions $d\bra S \ket/d\xi$ vs.~$\xi$. 
In the inset of Fig.~\ref{Fig8} we report $\Delta$ as a function of $m/n$. From this figure,
we observe a clear increase of $\Delta$ when decreasing the ratio $m/n$, an increase that seems to saturate
for ratios as small as $m/n\sim 1/100$.

\begin{figure*}[ht!]
\centering
\includegraphics[width=0.8\textwidth]{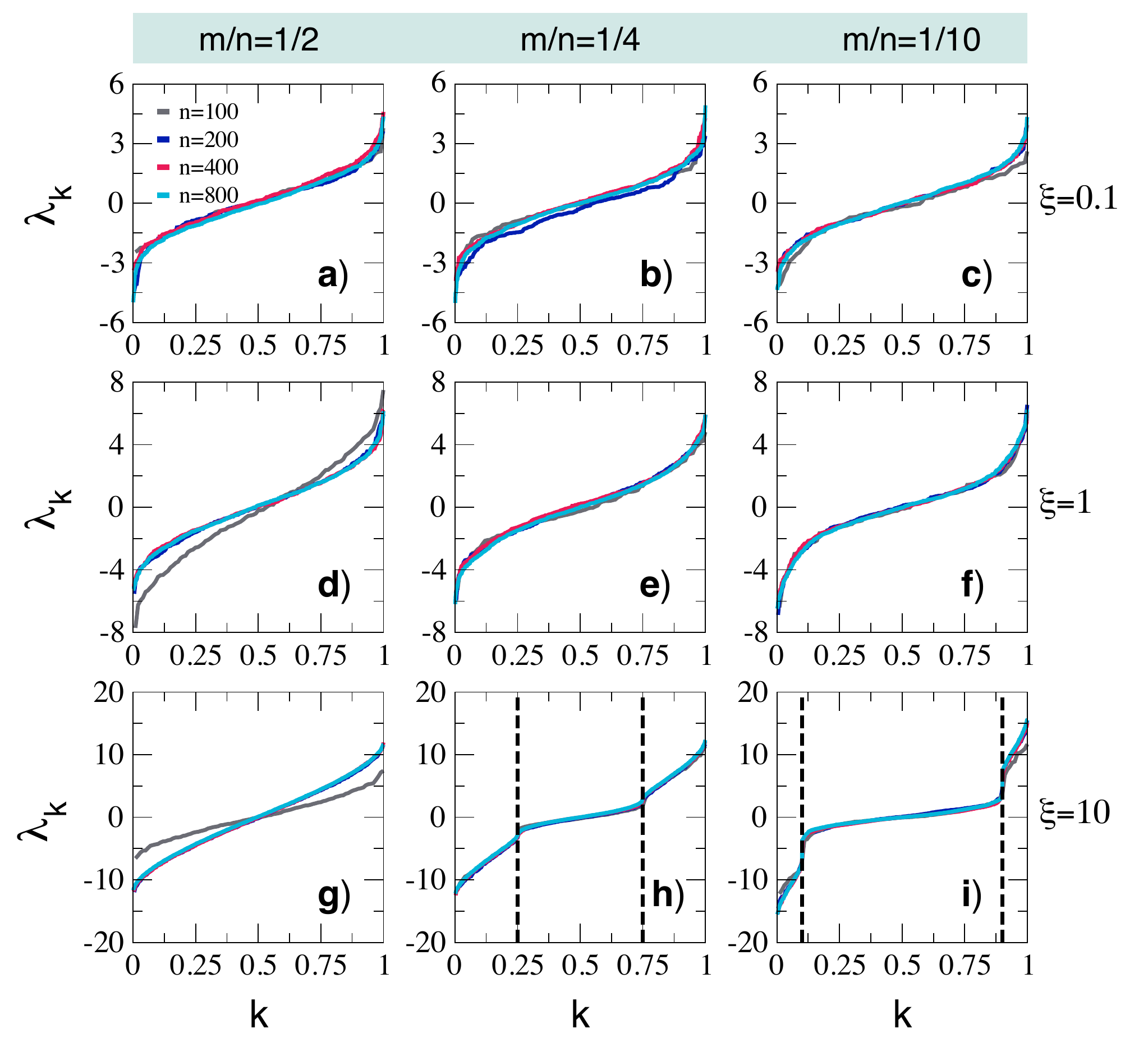} 
\caption{Eigenvalues $\lambda_k$ of the adjacency matrices of random bipartite graphs for several 
parameter combinations $(m,n,\alpha)$. Columns [rows] are characterized by a fixed $m/n$ [$\xi$].
A single graph realization is considered for each curve. Dashed lines in panels (h) and (i) coincide with those
in Figs.~\ref{Fig2}(b) and \ref{Fig2}(d), respectively.}
\label{Fig9}
\end{figure*}

It is worth stressing that once we have found that $\xi$ exists and that this parameter scales the eigenvector properties (characterized by their Shannon entropy) of the model of random bipartite graphs here studied, it is natural to expect that other properties (i.e.,~spectral properties, dynamical properties, transport properties, etc.) of the graph model would also scale with the same parameter. This is what we explore next, when we validate the previous surmise by closely inspecting the corresponding eigenvalues.

\section{Spectral properties}
\label{spectra}

In Fig.~\ref{Fig9}, we present the spectra of the adjacency matrices of random bipartite graphs 
for several combinations of the parameters $m$, $n$, and $\alpha$. Each panel is characterized by a fixed ratio $m/n$ 
and a fixed scaling parameter $\xi$. So, from the results in the previous Section, one should expect the
four spectra, reported in each of the panels of Fig.~\ref{Fig9} and corresponding to different graph sizes 
$n$, to fall one on top of the other. This is in fact the case, except for a small-size effect clearly observed 
in Fig.~\ref{Fig9}(d,g) when $n=100$. It is also interesting to note that the block structure of the adjacency 
matrix clearly reveals itself in the spectra, for large $\xi$ and small ratio $m/n$, see Fig.~\ref{Fig9}(h-i).

To characterize the spectral properties of the random bipartite graph model, we use the 
ratios of consecutive energy-level spacings $r$, which are defined as follows. Let $\{ \lambda \}$ be a set of ordered eigenvalues, the corresponding spacings $s_k$ are 
\begin{equation}
s_k = \frac{\lambda_{k+1}-\lambda_k}{\bra \lambda \ket} \ ,
\label{r}
\end{equation}
where $\bra \lambda \ket$ is the local mean eigenvalue density, while the ratios $r_k$ are defined 
as~\cite{OH07}
\begin{equation}
r_k = \frac{\mbox{min}(s_k,s_{k-1})}{\mbox{max}(s_k,s_{k-1})} \ ,
\label{r}
\end{equation}
such that $r_k\in[0,1]$ $\forall k$. Moreover, the probability distribution function of $r$ in the Poisson limit (which is reproduced by
our random bipartite graph model when $\alpha=0$) is~\cite{ABGR13}
\begin{equation}
P_{\mbox{\tiny P}}(r) = \frac{2}{(1+r)^2} \ .
\label{Prp}
\end{equation} 
Another important limit, that we will use as a reference, is the GOE case for which $P(r)$ 
gets the form~\cite{ABGR13}
\begin{equation}
\label{Prg}
P_{\mbox{\tiny GOE}}(r) = \frac{27}{4} \frac{r+r^2}{(1+r+r^2)^{5/2}} \ .
\end{equation}

It is important to stress that the nearest-neighbor energy-level spacing distribution $P(s)$~\cite{metha} 
is already a well accepted quantity to measure the degree of {\it chaos} or disorder in complex systems
and has been extensively used to characterize spectral properties of complex networks (see 
examples in~\cite{MAM15,MFR17b,AMG18} and references therein).
However, the use of $P(r)$ is more convenient here since it does not require the process known in RMT
as spectral unfolding~\cite{metha}, whose implementation for spectra with kinks as those in Figs.~\ref{Fig9}(h-i)
could be cumbersome.

\begin{figure*}[ht!]
\centering
\includegraphics[width=0.8\textwidth]{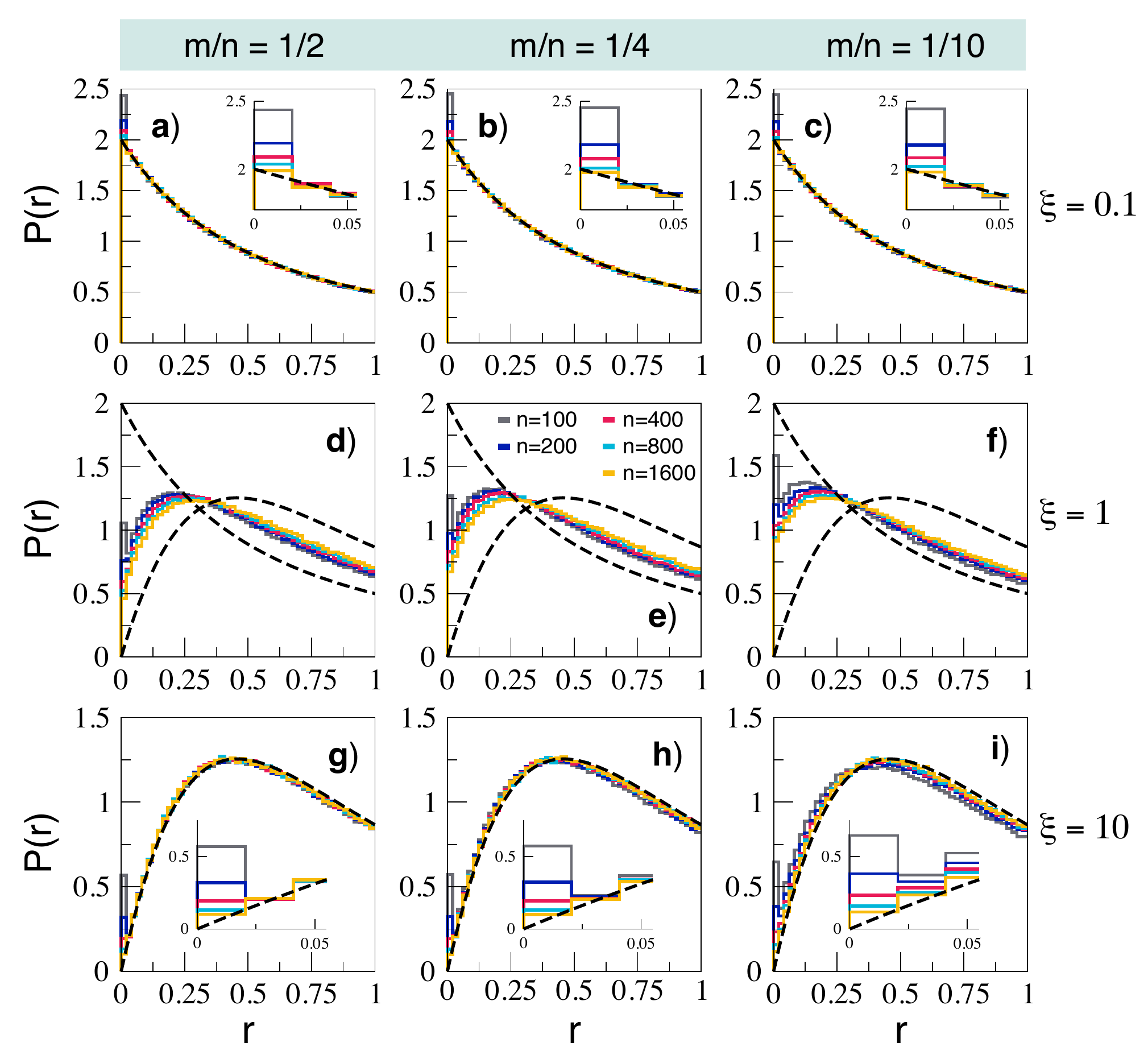} 
\caption{Distribution of ratios of consecutive energy-level spacings $P(r)$ for the eigenvalues of the adjacency 
matrices of random bipartite graphs with several parameters combinations $(m,n,\alpha)$. Columns [rows] 
are characterized by a fixed $m/n$ [$\xi$]. Each histogram is constructed with $10^6$ ratios.
Dashed lines in panels (a-c) [(g-i)] correspond to the RMT prediction for $P(r)$ in the Poisson [GOE] limit, 
see Eq.~(\ref{Prp}) [Eq.~(\ref{Prg})]. In panels (d-f) both equations, Eq.~(\ref{Prp}) and~(\ref{Prg}), are 
shown in dashed lines. Insets are enlargements of the main panels for $r$ close to zero.}
\label{Fig10}
\end{figure*}

Figure~\ref{Fig10} presents histograms of $P(r)$ for random bipartite graphs with several combinations of 
parameters $(m,n,\alpha)$. As well as in Fig.~\ref{Fig9}, each panel is characterized by a fixed 
ratio $m/n$ and a fixed scaling parameter $\xi$. 
With this figure we verify the invariance of $P(r)$ for fixed $\xi$, except for a small size effect that is enhanced 
at $r\to 0$; see the insets in panels (a-c,g-i) where the convergence to a steady $P(r)$ is obtained for 
large enough $n$. Besides, from Fig.~\ref{Fig10}, we
observe the Poisson to GOE transition in the shape of $P(r)$ when increasing $\xi$. Also, at the
transition borders, i.e.~at $\xi=0.1$ and $\xi=10$, the shape of $P(r)$ is well described by the corresponding
RMT predictions in the Poisson and GOE limits, respectively. This confirms our definition of the 
localization--to--delocalization transition region: $0.1<\xi<10$.
While, as expected, for intermediate values of 
$\xi$, see e.g.,~Fig.~\ref{Fig10}(d-f), $P(r)$ has a shape which is intermediate between $P_{\mbox{\tiny P}}(r)$ 
and $P_{\mbox{\tiny GOE}}(r)$. 

Finally, we would like to add that it is quite surprising that even for $m/n=1/10$ the $P(r)$ is very close to 
$P_{\mbox{\tiny GOE}}(r)$ when $\xi$ is large, see Fig.~\ref{Fig10}(i). Recall that for any $m/n<2$ the 
corresponding adjacency matrices have more null than not null off-diagonal matrix elements (see Fig.~\ref{Fig1}), 
therefore, being very different from members of the GOE. Moreover, we would also like to recall that we found 
that $S_{\tbox{MAX}}\approx \bra S \ket_{\tbox{GOE}}$ only for $m/n=1/2$, while 
$S_{\tbox{MAX}}<\bra S \ket_{\tbox{GOE}}$ for any $m/n<1/2$. Therefore, for our random bipartite graph model,
we can claim that $P(r)$ is less sensitive to deviations from GOE statistics than $\bra S \ket$.

\section{Conclusions}
\label{conclusions}

In this paper we have numerically studied the properties related to the eigenvectors and eigenvalues of the adjacency matrices 
of random bipartite graphs. Specifically, we have considered random bipartite graphs with self-loops, where all non-vanishing adjacency matrix 
elements are Gaussian random variables. 
Our random bipartite graph model depends on three parameters: The graph size $n$, the graph connectivity 
$\alpha$, and the size of the smaller set $m$ composing the bipartite graph.

First, through a proper scaling analysis of the Shannon entropy of the eigenvectors of the adjacency matrices
of such a random bipartite graph model, we defined a scaling parameter $\xi\equiv\xi(n,m,\alpha)$ that fixes the 
localization properties of the eigenvectors for a given ratio $m/n$.
Moreover, our analysis provides a way to predict the localization properties of the random bipartite graphs: 
For $\xi<0.1$ the eigenvectors are localized, the localization--to--delocalization transition occurs for 
$0.1<\xi<10$, whereas when $10<\xi$ the eigenvectors are extended. Next, to broaden the applicability of our findings, we demonstrated that for a fixed $\xi$, the spectral properties (characterized by the distribution of ratios of consecutive energy-level spacings) of the graph 
model are also universal, namely, they do not depend on the specific values of the bipartite graph parameters.

The results here derived are important in at least one applied field of research. Admittedly, the study of the stability of ecological systems makes use of the two main ingredients of our study. On the one hand, many ecosystems, including prey-predator and mutualistic systems, are faithfully represented by bipartite graphs, which are assumed to be random matrices when no information about the real structure is known. On the other hand, the analysis of the stability of such systems is often reduced to understand the eigenvalues and eigenvectors structure of the interaction matrices (or their Jacobian). Our results are important in so far they show that there are universal properties in such random bipartite networks, which might help to understand, in its turn, robust dynamical patterns of such systems regardless of their specific details such as size and interaction strengths. We plan to explore in more detail this potential application in the near future. 


\section*{Acknowledgements}

JAM-B acknowledges financial support from FAPESP (Grant No.~2019/ 06931-2), Brazil, and VIEP-BUAP (Grant No.~100405811-VIEP2019) and PRODEP-SEP (Grant No.~511-6/2019.-11821), Mexico. YM acknowledges partial support from the Government of Aragon, Spain through grant E36-17R (FENOL), by MINECO and FEDER funds (FIS2017-87519-P) and by Intesa Sanpaolo Innovation Center. The funders had no role in study design, data collection, and analysis, decision to publish, or preparation of the manuscript. 



\end{document}